\def\be{\begin{equation}}
\def\ee{\end{equation}}
\def\bea{\begin{eqnarray}}
\def\eea{\end{eqnarray}}

\documentclass[aps,pra,twocolumn,amsmath,amssymb]{revtex4}
\usepackage{epsfig,graphicx}
\usepackage{bm}

\begin{document}


\title{Quantum Bayesian approach to circuit QED measurement}

\author{Alexander N. Korotkov}
\affiliation{Department of Electrical Engineering, University of
California, Riverside, CA 92521-0204, USA}

\date{\today}


\begin{abstract}
We present a simple formalism describing evolution of a qubit in the
process of its measurement in a circuit QED setup. When a
phase-sensitive amplifier is used, the evolution depends on only one
output quadrature, and the formalism is the same as for a broadband
setup. When a phase-preserving amplifier is used, the qubit
evolution depends on two output quadratures. In both cases a perfect
monitoring of the qubit state and therefore a perfect quantum
feedback is possible.
\end{abstract}

\maketitle

\section{Introduction and qualitative discussion}\label{introduction}

The goal of this Lecture is to present a physical picture of the
process of continuous quantum measurement of a qubit in the circuit
quantum electrodynamics (cQED) setup
\cite{Wallraff-04,DiCarlo-10,Fragner-2008,Bertet,Siddiqi-2011}
(Fig.\ 1), extending or reformulating the previous theoretical
descriptions \cite{Blais-04,Clerk-RMP,Gambetta-06,Gambetta-08}.
Understanding of the qubit evolution in the process of measurement
is important for developing an intuition, which is useful in many
cases, in particular in designing various schemes of the quantum
feedback \cite{Wiseman-feedback,Hofmann-1998,Kor-feedback}.
   When a quantum measurement is discussed \cite{Braginsky}, there
are usually two different types of questions to answer: we can
either focus on obtaining information on the initial state (before
measurement) or focus on the quantum state after the measurement
(i.e.\ evolution in the process of measurement). Let us emphasize
that we consider the latter problem here and essentially extend the
collapse postulate by describing continuous evolution ``inside'' the
collapse timescale.

    In the cQED setup (Fig.\ 1) a qubit interacts with a GHz-range
microwave resonator, whose frequency slightly changes depending on
whether the qubit is in the state $|0\rangle$ or $|1\rangle$
\cite{Wallraff-04,DiCarlo-10,Fragner-2008,Bertet,Siddiqi-2011,Blais-04,Clerk-RMP,Gambetta-06,Gambetta-08}.
In turn, this frequency shift affects the phase (and in general
amplitude) of a probing microwave, which is transmitted through the
resonator (in another setup the microwave is reflected from the
resonator, but the difference is not important). The outgoing
microwave is amplified, and after that the rf signal is
downconverted by mixing it with the original microwave tone, so that
the low-frequency ($< 100$ MHz) output of the IQ mixer provides
information on the qubit state.  The output noise is mainly
determined by the first amplifying stage, the pre-amplifier. With
recent development of nearly quantum-limited superconducting
parametric amplifiers \cite{Devoret-ampl,Lehnert-ampl,Siddiqi-2011},
it is natural to use them as pre-amplifiers \cite{Siddiqi-2011}
instead of cryogenic high-electron-mobility transistors (HEMTs)
\cite{Wallraff-04,DiCarlo-10,Fragner-2008,Bertet}, which usually
have noise temperature above 3 K.

\begin{figure}[tb]
  \centering
\includegraphics[width=8.5cm]{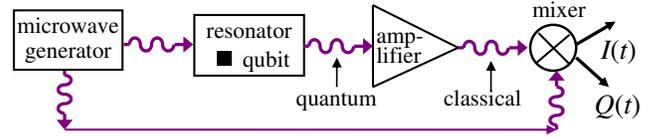}
  \caption{Schematic of the cQED setup. Microwave field of frequency
$\omega_m$ is transmitted through (or reflected from) the resonator
of frequency $\omega_r$, which slightly changes, $\omega_r\pm\chi$,
depending on the qubit state. After amplification the microwave is
sent to the IQ mixer, which produces two quadrature signals: $I(t)$
and $Q(t)$. For a phase-preserving amplifier we define $I(t)$ as the
quadrature carrying information on the qubit state, while for a
phase-sensitive amplifier we define $I(t)$ as corresponding to the
amplified quadrature.}
  \label{fig1}
\end{figure}

    Continuous quantum measurement in the cQED setup in some sense
falls in between a qubit measurement by a quantum point contact
(QPC) or a single-electron transistor (SET), the theory of which has
been developed over a decade ago \cite{Kor-99-01,Goan}, and
continuous quantum measurement in optics, for which the theory of
quantum trajectories has been developed even earlier
\cite{quant-traj}. Nevertheless, the cQED setup differs from both
these cases, and this is probably the reason why there is still a
confusion about the proper physical description of the measurement
process. The measurement by the QPC or SET is of the broadband type,
meaning that the monitored frequency band starts from zero. In
contrast, the cQED setup is of the narrowband type: we deal only
with a relatively narrow band around the probing microwave frequency
$\omega_m$. This necessarily involves two orthogonal quadratures
\cite{quadrature}: we work with rf signals of the type $A(t)\cos
(\omega_m t)+B(t)\sin(\omega_m t)$, and there are essentially two
signals $A(t)$ and $B(t)$ instead of only one in the broadband case.
In this sense the cQED setup is similar to the optical (especially
cavity QED) setup \cite{quant-traj}; however, there is an important
difference: in the cQED case the outgoing microwave is amplified
(Fig.\ 1) before being mixed with the original microwave, while
there is no amplification stage in the standard optical setup. The
operation will obviously depend on whether a phase-sensitive or a
phase-preserving amplifier is used, since a phase-preserving
amplifier necessarily adds the half quantum of noise into any
quadrature \cite{Haus-1962,Caves-1982,Dev-Likh,Clerk-RMP}. Notice
that the quantum trajectory theory for the cQED setup was developed
in \cite{Gambetta-08}; however, the amplifier stage was essentially
missing in the analyzed model.

    In this Lecture we consider the simplest cQED case, assuming
dispersive regime \cite{Blais-04}, exactly resonant microwave
frequency, absence of the Rabi drive, and sufficiently wide
resonator and amplifier bandwidths for the Markov approximation.
Some generalizations are rather straightforward; however, our goal
is a simple picture in a simple case.

  A description of continuous qubit measurement is essentially a
description of the quantum back-action. Following the same quantum
Bayesian framework as for the measurement by QPC/SET
\cite{Kor-99-01} (see \cite{Kor-rev} for review), we will discuss
two kinds of measurement back-action onto the qubit, which we name
here as ``spooky'' and ``realistic''. {\it The ``spooky'' (or
``quantum'', ``informational'', ``non-unitary'') back-action does
not have a physical mechanism} and therefore cannot be described by
the Schr\"odinger equation (in contrast to what people often think,
trying to find a mechanism for the quantum collapse); however, {\it
it is a common-sense consequence of acquiring information} on the
qubit state in the process of measurement. This is essentially the
same back-action which is discussed in the EPR paradox \cite{EPR}
and Bell inequality violation \cite{Aspect}; the only difference is
that in our case the information is incomplete and therefore we have
to use the quantum Bayes rule \cite{Caves-Bayes,Gardiner,Kor-rev}
instead of the projective collapse rule. In contrast, the
``realistic'' (or ``classical'', ``unitary'') back-action has a
physical mechanism: in the cQED case it is a fluctuation of the
number of photons in the resonator, which affects the phase of the
qubit state. The ``realistic'' back-action is usually discussed in
the standard theories of the cQED measurement
\cite{Blais-04,Clerk-RMP,Gambetta-06}. Actually, there is a certain
spookiness even in the ``realistic'' back-action (it may be affected
by a delayed choice, as discussed in Conclusion); however, we do not
want to emphasize it to keep the picture simple.
    When we measure the $z$-coordinate of the qubit state on the Bloch
sphere (the basis states $|1\rangle$ and $|0\rangle$ correspond to
the North and South poles), then the ``spooky'' back-action changes
the $z$-coordinate and leads to the state evolution along the
meridian lines, while the ``realistic'' back-action leads to the
evolution around the $z$-axis, i.e.\ along the parallels.

  It is important to notice that when the probing microwave leaves
the resonator after interaction with the qubit, one quadrature of
the microwave carries information about the qubit state, while the
orthogonal quadrature carries information on the fluctuating number
of photons in the resonator
\cite{Blais-04,Clerk-RMP,Gambetta-06,Gambetta-08}. Therefore, if a
phase-preserving amplifier is used, then the ``spooky'' and
``realistic'' back-actions are fully separated and correspond to two
orthogonal quadratures $I(t)$ and $Q(t)$ measured after the mixer
(it is trivial to choose the proper linear combinations of the I/Q
mixer outputs). The signals $I(t)$ and $Q(t)$ are necessarily noisy,
and the measurement back-actions are stochastic; however, there is a
correlation (full correlation in the ideal case) between the output
noise and the back-action noise in both channels. As a result
(derived later), for a quantum-limited phase-preserving amplifier
and absence of extra decoherence, the measured quadratures $I(t)$
and $Q(t)$ give us {\it full information about the back-action}, so
that a {\it random evolution of the qubit wavefunction can be
monitored precisely} (a useful analogy is with a Brownian particle
under a microscope: we cannot predict its motion, but we can monitor
it). This is what is needed, in particular, for arranging a perfect
quantum feedback control of the qubit state. It is interesting to
notice that for an ensemble-averaged evolution (in which the random
but monitorable qubit evolution is replaced by dephasing), exactly
one half of the ensemble dephasing $\Gamma$ comes from the
``spooky'' back-action, and the other half comes from the
``realistic'' back-action.

    In the case of a phase-sensitive amplifier it is sufficient to
measure after the mixer only the quadrature which was amplified; let
us still denote it as $I(t)$, though now its phase is determined by
the amplifier instead of the microwave-qubit interaction. In this
case the ``spooky'' and ``realistic'' back-actions are in general
mixed (not separated), because there is only one output signal
$I(t)$. This situation exactly corresponds to the broadband
measurement by the QPC/SET with a correlation between the output and
``realistic'' back-action noises \cite{Kor-rev}. The situation
simplifies when the amplified quadrature is the one which carries
information about the qubit state ($z$-coordinate). Then in the
quantum-limited case the ``realistic'' back-action is fully absent:
we cannot measure the photon number fluctuation and correspondingly
it does not fluctuate (in the imperfect case the effect of the
remaining ``realistic'' back-action can be described by an extra
dephasing). So we are left with only the ``spooky'' back-action, and
the quantum measurement description coincides with the simpler
theory of measurement by a symmetric QPC \cite{Kor-rev}, which does
not produce the ``realistic'' back-action. In contrast, in the case
when the photon-number quadrature is amplified, we do not obtain any
information on the qubit $z$-coordinate, and therefore there is no
``spooky'' back-action, but only the ``realistic'' one. In a general
case, when the amplified quadrature makes an arbitrary angle
$\varphi$ with the qubit-information quadrature, both types of the
back-action are present, and their strength depends on $\varphi$. It
is important to mention that the ensemble dephasing rate $\Gamma$
does not depend on $\varphi$, as required by causality. In
particular, in the quantum-limited case the contribution $\Gamma
\cos^2\varphi$ comes from the ``spooky'' back-action, while
$\Gamma\sin^2\varphi$ comes from the ``realistic'' back-action.

    Let us emphasize that both the phase-sensitive and
phase-preserving amplifiers permit exact monitoring of the qubit
state and therefore a perfect quantum feedback. The necessary
condition in both cases is that the detection system is
quantum-limited.

     In the following sections a formal description of
the above discussed results is presented. We start with reviewing
the Bayesian approach for the broadband qubit measurement, then
briefly discuss the difference between phase-preserving and
phase-sensitive amplifiers, and then present the formalism of the
narrowband continuous measurement of a qubit in the cQED setup. In
Conclusion we briefly discuss generalizations of the formalism,
quantum feedback, and the causality principle. We note that our
approach can be converted into the formal language of the
positive-operator-valued-measure(POVM)-type generalized quantum
measurement \cite{N-C} (then separation of the ``spooky'' and
``realistic'' back-actions corresponds to the decomposition of the
measurement operator into diagonal and unitary parts -- see later),
and our results for the case of a phase-sensitive amplifier are very
similar to the results of Ref.\ \cite{Gambetta-08}.

    \section{Broadband measurement}\label{broadband}

    In this section we review the Bayesian formalism
\cite{Kor-99-01,Kor-rev} for the broadband measurement of a qubit,
considering only the simple case without additional evolution, and
thus emphasizing the main physical idea of the formalism. We start
with the broadband formalism because it is simpler than for the
narrowband (cQED) measurement and it can be used as a natural step
in understanding the cQED setup.

    For definiteness let us assume that the qubit is a double quantum
dot populated with one electron (Fig.\ 2), and the states
$|0\rangle$ and $|1\rangle$ correspond to the electron localized in
one or the other dot. The qubit is measured by a small-transparency
tunnel junction (model of QPC), whose barrier height depends on the
electron location, so that the two qubit states correspond to
different average currents $I_0$ and $I_1$ through the QPC. The
voltage across the QPC is sufficiently large to make the detector
output classical (Markov approximation), and $|\Delta I|\ll |I_c|$,
where $\Delta I=I_1-I_0$ is the response and $I_c=(I_0+I_1)/2$ is
the mean value; this weak response assumption allows us to consider
the QPC current $I(t)$ as a quasicontinuous noisy signal (see
\cite{Kor-rev} for the detailed discussion of required assumptions;
the formalism needs only a minor change if $\Delta I\sim I_c$).
    Then the output signal of the detector is
    \be
    I(t)=I_c + (\Delta I/2) \, z(t) +\xi (t), \,\,\,
    S_\xi (\omega ) =S,
    \ee
where $z=\rho_{11}-\rho_{00}$ is the $z$-component of the Bloch
sphere representation of the qubit density matrix $\rho(t)$, and
$\xi (t)$ is the white shot noise with spectral density $S=2eI_c$
(we use the single-sided definition for the spectral density, in
which the signal variance (``power'') corresponds to $\int_0^\infty
S(\omega)\, d\omega/2\pi$; the definition of $S$ is twice smaller in
Ref.\ \cite{Clerk-RMP} and $4\pi$ times smaller in Refs.\
\cite{Gambetta-06,Gardiner}). We emphasize that the detector signal
$I(t)$ is classical, and the qubit state $\rho(t)$ is practically
unentangled from the detector, but obviously depends on $I(t)$.

\begin{figure}[tb]
  \centering
\includegraphics[width=5.0cm]{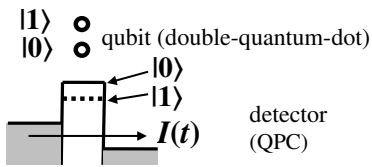}
  \caption{Schematic of a broadband measurement setup:
double-quantum-dot qubit is measured by a QPC (tunnel junction). The
output signal $I(t)$ is the QPC current. }
  \label{fig2}
\end{figure}

The detector Hamiltonian and the qubit-detector interaction
Hamiltonian are given in Refs.\ \cite{Kor-99-01,Kor-rev}, they are
not really important for our discussion here. For simplicity let us
assume that the qubit Hamiltonian is zero, $H_{qb}=0$, so that the
qubit evolution is due to the measurement only. In this case the
qubit evolution during time $t$ happens to be determined only by the
time-averaged value of the measured detector output
    \be
    \overline{I}_m (t) =\frac{1}{t} \int_0^t I(t')\, dt' ,
    \ee
which would contain full information for a classical measurement.

  Because of the correspondence principle, the evolution of the
diagonal elements of the qubit density matrix $\rho$ (${\rm Tr}
\rho=1$) should correspond to the classical evolution of
probabilities, which are given by the classical Bayes rule. The
Bayes rule says that an updated (a posteriori) probability of a
system state is proportional to the initial (a priori) probability
and the probability (likelihood) of the obtained measurement result
assuming this particular state. In our case $\overline{I}_m(t)$ is
the measurement result, and its probability for the qubit in the
basis state $|j\rangle$ has the Gaussian distribution
    \be
P_{|j\rangle}(\overline{I}_m)=\frac{1}{\sqrt{2\pi D}} \, \exp
[-(\overline{I}_m-I_j)^2/2D], \,\, D=\frac{S}{2t},
    \label{gaussian}\ee
 where $D$ is the variance, which
decreases with the measurement time $t$. Therefore the
correspondence principle demands the Bayesian evolution
    \be
    \frac{\rho_{11}(t)}{\rho_{00}(t)} =
    \frac{\rho_{11}(0)}{\rho_{00}(0)} \, \frac{\exp
[-(\overline{I}_m(t)-I_1)^2/2D]}{\exp
[-(\overline{I}_m(t)-I_0)^2/2D]},
    \label{rho-diag}\ee
which in our terminology is due to the ``spooky'' back-action; it
cannot be described by the Schr\"odinger equation, but follows from
common sense.

    If the phase of qubit state is not affected by the
measurement process (no ``realistic'' back-action), then an
arbitrary initial wavefunction $|\psi(0)\rangle=
\sqrt{\rho_{00}(0)}\, |0\rangle +e^{i\phi}\sqrt{\rho_{11}(0)}\,
|1\rangle$ becomes $|\psi(t)\rangle =  \sqrt{\rho_{00}(t)}\,
|0\rangle +e^{i\phi}\sqrt{\rho_{11}(t)}\, |1\rangle$ with the same
phase $\phi$; therefore for an arbitrary mixed state we get
    \be
 \rho_{01}(t)= \rho_{01}(0)\, \frac{\sqrt{\rho_{00}(t)\,
\rho_{11}(t)}} {\sqrt{\rho_{00}(0)\, \rho_{11}(0)}}\, .
    \label{rho-off}\ee
Equations (\ref{rho-diag}) and (\ref{rho-off}) describe the
``spooky'' back-action.

    Now assume that due to the qubit-detector interaction (e.g.\
Coulomb interaction), each electron passing through the detector
rotates the qubit phase $\phi$ by a small amount $\Delta\phi$. From
the measured result $\overline{I}_m(t)$ we know exactly how many
electrons passed through [$n_e=\overline{I}_m(t)t/q$ with $q$ being
the electron charge], and can easily introduce the corresponding
phase factor into Eq.\ (\ref{rho-off}):
        \be
 \rho_{01}(t)= \rho_{01}(0)\, \frac{\sqrt{\rho_{00}(t)\,
\rho_{11}(t)}} {\sqrt{\rho_{00}(0)\, \rho_{11}(0)}}\, \exp[iK
\overline{I}_m(t)\, t] ,
    \label{rho-off-2}\ee
where $K=\Delta\phi/q$. The non-stochastic factor $\exp(iKI_ct)$ can
be obviously ascribed to the qubit Hamiltonian; however, this is not
important here. The factor $\exp[iK \overline{I}_m(t)\, t]$ in Eq.\
(\ref{rho-off-2}) is the effect of the ``realistic'' back-action. It
may or may not be present in a particular physical situation; for
example, $K=0$ for measurement by a symmetric QPC, while $K\neq 0$
in an asymmetric QPC or SET case.

    Finally, if there is an extra pure dephasing of a qubit with
rate $\gamma$, then Eq.\ (\ref{rho-off-2}) becomes
        \be
 \rho_{01}(t)= \rho_{01}(0)\, \frac{\sqrt{\rho_{00}(t)\,
\rho_{11}(t)}} {\sqrt{\rho_{00}(0)\, \rho_{11}(0)}}\, \exp[iK
\overline{I}_m(t)\, t] \, e^{-\gamma t} .
    \label{rho-off-3}\ee
Equations (\ref{rho-diag}) and (\ref{rho-off-3}) is {\it the main
starting point of the Bayesian formalism} \cite{Kor-rev}. It is then
easy to include non-zero qubit Hamiltonian $H_{qb}$ by
differentiating Eqs.\ (\ref{rho-diag}) and (\ref{rho-off-3}) over
time (paying attention to whether the Stratonovich or It\^o
definition of the derivative is used) and adding terms due to
$H_{qb}$. Energy relaxation and other mechanisms of the qubit state
evolution can be included in the same way.
   Actually, there are many ways to derive the Bayesian equations
(\ref{rho-diag}) and (\ref{rho-off-3})
\cite{Kor-99-01,Kor-rev,Goan,Jordan,Nazarov}, but we focus here only
on their meaning, not on their derivation.

    Notice that averaging of Eqs.\ (\ref{rho-diag}) and
(\ref{rho-off-3}) over the measurement result $\overline{I}_m$
(i.e.\ ensemble averaging) with the probability distribution
    \be
    P(\overline{I}_m)=\rho_{00}(0)P_{|0\rangle}(\overline{I}_m)+
\rho_{11}(0)P_{|1\rangle}(\overline{I}_m)
    \label{P(I)}\ee
 gives the same evolution as for a
pure dephasing: the diagonal matrix elements of $\rho$ do not
evolve, while the off-diagonal element $\rho_{01}$ decays as
$\rho_{01}(0)\, e^{-\Gamma t}$ with the ensemble dephasing rate
\cite{Kor-rev}
    \be
    \Gamma = \frac{(\Delta I)^2}{4S}+ \frac{K^2S}{4}+\gamma ,
    \label{Gamma}\ee
which has clear contributions from the ``spooky'' back-action,
``realistic'' back-action, and additional dephasing.

    In the case $\gamma=0$ an initially pure qubit state remains
pure; in other words we can monitor evolution of a qubit
wavefunction. This property can be used as the definition of a
quantum-limited detector \cite{Kor-99-01,Kor-rev}. The quantum
efficiency $\eta$ can then be naturally defined as $\eta
=1-\gamma/\Gamma$. If by some reason the ``realistic'' back-action
is considered as dephasing (i.e.\ only in the averaged way), then
the quantum efficiency can be defined as $\tilde\eta=1-\gamma/\Gamma
- K^2S/4\Gamma$ (here the definitions of $\tilde\eta$ and $\eta$ are
exchanged compared with the definitions in \cite{Kor-rev}). In other
words, $\tilde\eta =(\Delta I)^2/4S\Gamma$ is the relative
contribution of only the ``spooky'' back-action in the ensemble
dephasing $\Gamma$. In particular, this definition is relevant to
the peak-to-pedestal ratio of the Rabi spectral peak \cite{Kor-osc},
which is equal to $4\tilde\eta$.
 As an example, if
$\gamma=0$ and contributions in Eq.\ (\ref{Gamma}) from the
``spooky'' and ``realistic'' back-actions are equal to each other
(as in the cQED setup with a phase-preserving amplifier), then
$\eta=1$ but $\tilde{\eta}=1/2$.

    A non-ideal detector ($\eta<1$) can be modeled in two equivalent
ways \cite{Kor-nonideal}: we either add an extra dephasing $\gamma$
to the qubit or we add an extra noise to the output of the ideal
detector. Only the total dephasing $\Gamma$, response $\Delta I$,
total output noise $S$, and correlation factor $K=
\delta\langle\phi\rangle/\delta(\overline{I}_m t)$ are the physical
(i.e.\ experimentally measurable) parameters, while distribution of
the non-ideality between the extra dephasing and additional output
noise is a matter of convenience [here $\phi={\rm arg(\rho_{01})}$,
and notation $\langle \phi\rangle$ reminds about averaging over
additional classical noise at the output]. We emphasize that the
Bayesian formalism deals only with the experimentally measurable
parameters $\Delta I$, $S$, $K$, $\Gamma$, and the output signal
$I(t)$.

    In the ideal case ($\eta=1$) the evolution equations
(\ref{rho-diag}) and (\ref{rho-off-2}) can be translated into the
language of POVM-type generalized measurement. In this approach the
effect of measurement is described as  \cite{N-C}
    \be
  |\psi (t)\rangle =\frac{M_R|\psi(0)\rangle}{||M_R|\psi(0)\rangle ||},
   \,\,\,
  \rho (t)=\frac{M_R\rho(0)M_R^\dagger}{\mbox{Tr}[M_R^\dagger
  M_R\rho(0)]},
    \ee
where $M_R$ is the so-called measurement (Kraus) operator,
corresponding to the result $R$. The probability of the result $R$
is $P_R=||M_R|\psi(0)\rangle||^2$ using wavefunctions or $P_R=\mbox
{Tr} [M_R^\dagger M_R\,\rho(0)]$ using density matrices;  therefore
the POVM elements $M_R^\dagger M_R$ should satisfy the completeness
condition $\sum_R M_R^\dagger M_R = \openone$.
    The relation  between this
approach and the quantum Bayesian approach can be understood via the
operator  decomposition
    \be
    M_R = U_R \sqrt{M_R^\dagger M_R},
    \ee
where $U_R$ is  unitary and the square root of the positive operator
$M_R^\dagger M_R$ is defined in the natural way in the diagonalizing
basis. It is easy to see that $\sqrt{M_R^\dagger M_R}$ is
essentially the quantum Bayes rule (in the diagonalizing basis); in
our terminology it corresponds to the ``spooky'' back-action, while
$U_R$ corresponds to the ``realistic'' back-action.
    For the discussed setup the result $R$ is $\overline{I}_m(t)$, the
``spooky'' back-action $[M_R^\dagger M_R]^{1/2}$ should be
determined by the probabilities $P_{|j\rangle}(\overline{I}_m)$
given by Eq.\ (\ref{gaussian}), and the ``realistic'' back-action
$U_R$ is given by the phase factor in Eq.\ (\ref{rho-off-2}).
Therefore the corresponding measurement operator is
    \begin{eqnarray}
&&    M(\overline{I}_m) = \exp ( -i K \overline{I}_m t \,
\sigma_z/2)
    \nonumber \\
&&\hspace{0.5cm} \times \left[\sqrt{P_{|0\rangle}(\overline{I}_m)}
\, |0\rangle \langle 0| + \sqrt{P_{|1\rangle}(\overline{I}_m)} \,
|1\rangle \langle 1|\right], \qquad
    \end{eqnarray}
where $\sigma_z$ is the Pauli matrix.

    \section{Phase-preserving vs.\ phase-sensitive
    amplifiers}\label{amplifiers}

    Before discussing microwave amplifiers, let us consider a
measurement of an oscillator, for example, a mechanical resonator
with frequency $\omega_r$ and mass $m$. This is a very well studied
problem \cite{Clerk-RMP,Braginsky}, so we will only discuss a way to
understand the results. A classical resonator position $x$
oscillates as $x_c(t)=A\cos(\omega_r t)+B\sin(\omega_r t)$ (in this
section $x$ stands for the usual spatial coordinate, not for the
Bloch sphere coordinate). The corresponding quantum state is called
the ``coherent state'' in the optical language; it is represented by
the wavefunction $\psi (x,t)=\psi_{gr}[x-x_c(t)]\exp(i p_c
x/\hbar)$, where $\psi_{gr}(x)$ is the ground state and
$p_c=m\dot{x}_c(t)$ is the classical momentum. So the coherent state
is essentially the ground state with oscillating center position.
Notice that continuous quantum measurement of a resonator position
can be described in the same Bayesian way \cite{Ruskov-resonator} as
in the previous section; for example the ``spooky'' back-action
gives the evolution
$\psi(x,t)=\psi(x,0)\exp[-(\overline{I}_m(t)-I(x))^2/4D]/{\rm
Norm}$, where $I(x)$ is the average detector signal for the
resonator position $x$, and ${\rm Norm}$ is normalization [see Eqs.\
(\ref{rho-diag}) and (\ref{rho-off})]. The time step $t$ in this
case should be chosen much shorter than  $\omega_r^{-1}$ so that the
unitary evolution and evolution due to measurement may be simply
added.

Let us consider the following game: Charlie prepares an oscillator
in a coherent state with quadratures $A$ and $B$, gives it to David,
and David's goal is to find $A$ as accurately as possible. An
optimal strategy is rather obvious: David should make a projective
measurement of $x$ at time $t=2\pi n/\omega_r$ with any integer $n$
(to avoid contribution from the $B$-term), and the measurement
result is the best estimate of $A$ [if the measurement is done at
$t=(2\pi n+\pi)/\omega_r$, then the result should be multiplied by
$-1$]. Even though the strategy is optimal, the inaccuracy of
David's result is obviously the width (standard deviation)
$\sigma_{gr}=\sqrt{\hbar/2m\omega_r}$ of the ground state shape
$|\psi_{gr}(x)|^2$; in energy units this inaccuracy corresponds to
one half of the energy quantum.

    Now assume that David cannot make projective measurements, but
only ``finite-strength'' (i.e.\ imprecise) measurements. The best
accuracy $\sigma_{gr}$ can still be achieved if the measurement is
done in the simple but very clever ``quantum non-demolition'' (QND)
way: many finite-strength measurements are made at times $t=2\pi
n/\omega_r$; this is called ``stroboscopic'' measurement
\cite{Braginsky}. Since the oscillator returns to the same state
after the period $2\pi/\omega_r$, the unitary evolution is not
important, and many finite-strength measurements (described by the
Bayesian equation above) are ``stacked'' to produce a strong,
essentially projective measurement. More generally, the necessary
condition to have the best accuracy $\sigma_{gr}$ for $A$ is that
the measurement is not sensitive to the quadrature $B$.

    Now assume that David is only allowed to make a continuous
measurement with unmodulated weak strength (so that the inaccuracy
achieved after $\omega_r^{-1}$ is much larger than $\sigma_{gr}$).
Then the ``spooky'' back-action gets mixed with the unitary
evolution, essentially adding noise into the monitored evolution, so
that after a while the resonator state becomes mostly determined by
the back-action and almost not dependent on the initial state. As
the result, the best accuracy for measurement of $A$ becomes
$\sqrt{2}\, \sigma_{gr}$, which in energy units corresponds to two
half-quanta \cite{Braginsky}, that is twice worse than for the
projective or stroboscopic measurement. However, the continuous
monitoring gives us an information about $B$ in the same way as for
$A$, so the accuracy of $B$-measurement is also $\sqrt{2}\,
\sigma_{gr}$. Therefore, in some sense continuous phase-insensitive
measurement brings the same total information as the phase-sensitive
(e.g.\ stroboscopic) measurement; however, in our game only half of
this information is useful for the David's goal.

    After discussing measurement of the resonator state it is easy
to understand the quantum limits for the high-gain microwave
amplifiers. Now suppose Charlie prepares a coherent state of a
microwave resonator with quadratures $A$ and $B$, gives it to David
to find $A$, and David uses an amplifier for amplification of the
microwave field, which slowly leaks from the resonator until it is
empty. There is only
 classical signal processing after the amplifier, so amplification
is essentially the quantum measurement. The results are the same as
above \cite{Clerk-RMP,Braginsky,Caves-1982,Dev-Likh}: a
phase-sensitive amplifier, which amplifies only $A$-quadrature and
``de-amplifies'' (attenuates) $B$-quadrature can measure $A$ with
accuracy $\sigma_{gr}$, while a phase-preserving amplifier can
measure $A$ only with accuracy $\sqrt{2}\, \sigma_{gr}$ (and also
measures $B$ with the same accuracy $\sqrt{2}\, \sigma_{gr}$).
Technically, the accuracy is limited by the noise at the amplifier
output, so this noise should forbid measuring of $A$ with accuracy
better than $\sigma_{gr}$ by a phase-sensitive amplifier, and better
than $\sqrt{2}\, \sigma_{gr}$ by a phase-preserving amplifier.
Therefore in the quantum-limited case the output noise power of a
phase-preserving amplifier (per quadrature) is twice larger than for
a phase-sensitive amplifier with the same gain; this is often called
an ``additional noise'', corresponding to one half of the energy
quantum (one more half-quantum is present in both cases)
\cite{Clerk-RMP,Braginsky,Caves-1982,Dev-Likh}. It may be somewhat
confusing why this result does not depend on the rate with which the
microwave leaks from the resonator. So let us check the scaling: for
$k$ times slower leakage ($k$ times larger $Q$-factor) the microwave
amplitude is $\sqrt{k}$ times smaller, but accumulation time is $k$
times longer, therefore the measured signal for the quadrature $A$
is $\sqrt{k}$ times larger, which is the same factor as for the
noise accumulation. Therefore, the signal-to-noise ratio which
determines the $A$-accuracy does not depend on the leakage rate
(resonator bandwidth).

   \section{Narrowband (cQED) measurement}\label{narrowband}

    Using the above discussion of microwave amplifiers, it is easy
to extend the Bayesian approach for a broadband quantum measurement
to the narrowband cQED setup.

We consider the standard cQED setup
\cite{Wallraff-04,DiCarlo-10,Fragner-2008,Bertet,Siddiqi-2011,Blais-04,Clerk-RMP,Gambetta-06,Gambetta-08},
in which a qubit interacts with a microwave resonator, and assume
the dispersive regime with the Hamiltonian
   \begin{equation}
H=(\hbar\omega_{qb}/2)\,\sigma_z +\hbar \omega_r a^\dagger a +\hbar
\chi a^\dagger a \sigma_z ,
    \label{H-dispersive}\end{equation}
where $\omega_{qb}=\omega_{qb, bare}+\chi$ is the Lamb-shifted qubit
frequency with no photons in the resonator,
$\chi=g^2/(\omega_{qb,bare}-\omega_r)$ is the effective coupling
with $g$ being the Jaynes-Cummings coupling, $\omega_r$ is the bare
resonator frequency, Pauli operator $\sigma_z$ acts on the qubit
state in the energy basis $\{|0\rangle,|1\rangle\}$, and the
resonator creation/annihilation operators are $a^\dagger$ and $a$.
Notice that the resonator frequency increases by $2\chi$ when the
qubit state changes from $|0\rangle$ to $|1\rangle$; conversely, the
qubit frequency increases by $2\chi$ per each additional photon in
the resonator.
    To measure the qubit state, a microwave field with frequency
$\omega_m$ is either transmitted through or reflected from the
resonator, then amplified and sent to the IQ mixer, which measures
both quadratures relative to the original microwave tone (Fig.\ 1).
The qubit state affects the resonator frequency and therefore
affects the phase (and in general amplitude) of the outgoing
microwave.

    An elementary Fabry-P\'erot analysis gives the classical
(complex) microwave field $F_r$ inside the resonator:
    \be
    F_r =\frac{2 F_{in} t_{in}/\kappa\tau_{rt}}{1-2i(\omega_m-\omega_r)/
\kappa},
    \label{Fabry-Perot}\ee
where $F_{in}$ is the applied incident field, $t_{in}$ is the
transmission amplitude of the barrier from the incident side,
$\kappa$ is the resonator bandwidth due to the microwave leakage
from the both sides (the $Q$-factor is $\omega_r/\kappa$), and the
round-trip time is $\tau_{rt}= 2\pi/\omega_r$ for a half-wavelength
resonator and $\tau_{rt}= \pi/\omega_r$ for a quarter-wavelength
resonator. A similar formula with the same denominator describes a
lumped resonator. In presence of the qubit, the resonator frequency
$\omega_r$ in this formula is substituted by $\omega_r\pm \chi$,
depending on the qubit state. Notice that for the quantum
measurement analysis {\it there is no difference between the cases
of transmission and reflection} for the same $F_r$ and $\kappa$,
because the field leaking from the resonator is determined only by
$F_r$ and $\kappa$. (The reflection case has a technical advantage
of dealing with a twice smaller outgoing microwave field for the
same measured signal.) However, an important parameter is the
collected fraction $\eta_{col}=\kappa_{col}/\kappa$ of the leaking
microwave power; we will often assume the ideal case $\eta_{col}=1$
(for the transmission setup this requires strongly asymmetric
coupling, $|t_{in}|\ll |t_{out}|$).

    For simplicity we assume the resonant case, $\omega_m=\omega_r$,
then the ensemble qubit dephasing due to measurement is
\cite{Blais-04,Clerk-RMP}
    \be
    \Gamma = 8\chi^2\bar{n}/\kappa,
    \label{Gamma-cQED}\ee
where $\bar{n}$ is the average number of photons in the resonator.
It is easy to include Rabi oscillations into the model; however, we
do not do it for simplicity and also for more transparent analogy
with Sec.\ \ref{broadband}, in which we considered a qubit with zero
Hamiltonian, evolving only due to measurement; this case exactly
corresponds to the cQED Hamiltonian (\ref{H-dispersive}) in the
rotating frame.

    We will need several assumptions to describe the qubit state
evolution in the process of measurement. First, for the validity of
the dispersive approximation (\ref{H-dispersive}) we need
sufficiently large qubit-resonator detuning,
$|\omega_{qb}-\omega_r|\gg |g|$, and not too many photons in the
resonator, $\bar{n}\ll (\omega_{qb}-\omega_r)^2/g^2$ (we do not
consider the recently discovered nonlinear regime \cite{nonlinear}).
Second, to use the Markov approximation for the evolution we need
the so-called ``bad cavity'' assumption: $\Gamma \ll \kappa \ll
\omega_r$ (if the qubit evolves due to Rabi oscillations with
frequency $\Omega_R$, we also need $\Omega_R\ll \kappa$). This
assumption means that the photons leave the resonator much faster
than evolution of the qubit state, and therefore there is
practically no entanglement between the qubit and unmeasured
microwave field. This assumption also implies that the two resonator
states for the qubit states $|0\rangle$ and $|1\rangle$ are almost
indistinguishable, $\bar{n}(\chi/\kappa)^2\ll 1$. Third, we use the
``weak response'' assumption, which requires a small phase
difference between the two resonator states, $|\chi|/\kappa \ll 1$.
This means that each outgoing photon carries only a little
information about the qubit state. Notice that for $\bar{n}\agt 1$
the previous assumption $\kappa\gg\Gamma$ automatically implies the
weak response, and even for $\bar{n}\ll 1$ the weak response
assumption is not always needed. Fourth, we will neglect the qubit
energy relaxation due to measurement \cite{Blais-04,Clerk-RMP},
which can be added later.

    A coherent state in the resonator with average $\bar{n}$ photons
and zero average phase corresponds to the oscillation of the field
expectation value $\langle F_r(t)\rangle=2\sqrt{\bar{n}}\,
\sigma_{gr} \cos (\omega_m t)$, where $\sigma_{gr}$ is the ground
state width (rms uncertainty) and we assume $\omega_m=\omega_r$.
(Notice that the amplitude $\sigma_{gr}$ corresponds to 1/4 photon.)
Interaction with the qubit slightly changes the phase,
$\cos(\omega_mt \mp 2\chi/\kappa)$, depending on the qubit state, so
that
    \begin{eqnarray}
&& \langle F_r(t)\rangle = A\cos(\omega_m t)+B\sin (\omega_m
t),\,\,\,
 A=2\sqrt{\bar{n}} \,
    \sigma_{gr} , \quad
    \nonumber \\
&&
   B=\pm (4\chi /\kappa ) \sqrt{\bar{n}}\, \sigma_{gr} =
   (4\chi /\kappa ) \sqrt{\bar{n}}\, \sigma_{gr} \, z,
    \label{A-B-quadratures}\end{eqnarray}
 where $z$ is the qubit Bloch coordinate.
Thus the small $B$-quadrature carries information about the qubit
state, while larger $A$-quadrature may give us information on the
fluctuations of the photon number in the resonator. In the optical
representation (Fig.\ 3) with axes $A/\sigma_{gr}$ and
$B/\sigma_{gr}$, the two resonator states for the qubit states
$|0\rangle$ and $|1\rangle$ are shown as two ``error circles''
\cite{Walls-Milburn} with rms uncertainty 1 along any direction and
distance $2\sqrt{\bar{n}}$ between the origin and circle centers (if
axes $A/2\sigma_{gr}$ and $B/2\sigma_{gr}$ are used, then the
distance is $\sqrt{\bar{n}}$, while the uncertainty is 1/2).

\begin{figure}[tb]
  \centering
\includegraphics[width=5.0cm]{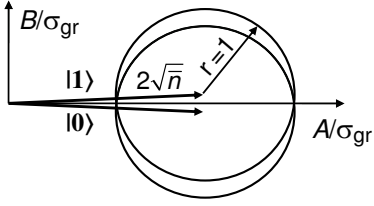}
  \caption{Phase space representation: for each qubit state the coherent
state with $\langle F_r\rangle=A\cos(\omega_rt)+B\sin(\omega_rt)$ in
the resonator is shown \cite{Walls-Milburn} as an ``error circle''
with radius 1, shifted by $2\sqrt{\bar{n}}$ from the origin. Axes
are normalized by the standard deviation $\sigma_{gr}$ of the ground
state. The $B$-quadrature carries information about the qubit state,
the $A$-quadrature corresponds to the number of photons in the
resonator.}
  \label{fig3}
\end{figure}

    \subsection{Phase-sensitive amplifier}

    Let us start with the case when a phase-sensitive amplifier is
used in the cQED setup. Also, we first assume the most ideal case:
the amplifier is quantum-limited, it amplifies the optimal
$B$-quadrature, there is no microwave collection loss
($\kappa_{col}/\kappa=1$), and there is no extra noise or dephasing.
Then, as discussed in the previous section, measuring once the
microwave contents of the resonator (by fully emptying it) we can
measure the $B$-quadrature with imprecision $\sigma_{gr}$.
Therefore, in the continuous measurement for time $t$ the
$B$-quadrature is measured with imprecision
$\sigma_{gr}/\sqrt{\kappa t}$, which converts into the imprecision
$\sqrt{\kappa/t}/(4\chi\sqrt{\bar{n}})$ of the qubit $z$-coordinate.
Following the language of Sec.\ \ref{broadband}, let us discuss the
signal and noise at the output of the setup. There are two outputs
of the IQ mixer; however, only the amplified quadrature carries an
information, so let us denote the corresponding output of the mixer
(or their linear combination) as $I(t)$. Then the response $\Delta
I=I_1-I_0$ corresponds to $\Delta z=2$ and $\Delta
B=8(\chi/\kappa)\sqrt{\bar{n}}\,\sigma_{gr}$. For measurement during
time $t$ the above variance $(\kappa/t)/(4\chi\sqrt{\bar{n}})^2$ of
the $z$-coordinate converts into the variance $(\kappa/t)(\Delta
I/8\chi \sqrt{\bar{n}})^2$ of the measured output
$\bar{I}_m=(1/t)\int_0^t I(t')\, dt'$. Equating it with $D=S/2t$, we
find the (single-sided) spectral density of the $I(t)$ noise:
    \be
    S_{min}=(\Delta I_{max})^2\kappa /(32 \chi^2 \bar{n}),
    \label{S}\ee
where we replaced $S$ with $S_{min}$ and $\Delta I$ with $\Delta
I_{max}$ to remind that we consider the quantum-limited case, and
the response is maximized by amplifying the optimal quadrature.
Notice that since $\Delta I_{max}\propto \chi
\sqrt{\bar{n}/\kappa}$, the noise $S_{min}$ does not depend on the
qubit or resonator properties; it is essentially the amplified
vacuum noise and depends only on the amplifier gain.

\begin{figure}[tb]
  \centering
\includegraphics[width=7cm]{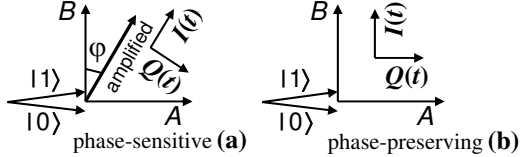}
  \caption{Relations between relevant quadratures. The quadratures
$A$ and $B$ are for the microwave field in the resonator. (a) For a
phase-sensitive amplifier the amplified quadrature makes an angle
$\varphi$ with the informational $B$-quadrature. The corresponding
quadrature at the mixer output is defined as $I(t)$ [the output
$Q(t)$ is then useless]. (b) For a phase-preserving amplifier we
define output quadratures $I(t)$ and $Q(t)$ as corresponding to the
resonator quadratures $B$ and $A$.}
  \label{fig4}
\end{figure}

    Obtaining information on the qubit $z$-coordinate via the signal $I(t)$
with response $\Delta I_{max}$ and noise $S_{min}$, we necessarily
cause the ``spooky'' back-action described by Eqs.\ (\ref{rho-diag})
and (\ref{rho-off}).  As discussed in Sec.\ \ref{broadband}, this is
the consequence of the corresponding principle or just the common
sense. Now averaging the $\rho_{01}$ evolution in Eq.\
(\ref{rho-diag}) over the measurement result $\bar{I}_m$ with its
probability distribution (\ref{P(I)}) and (\ref{gaussian}), we see
that the ``spooky'' back-action dephases an ensemble of qubits with
the rate [see Eq.\ (\ref{Gamma})] $(\Delta
I_{max})^2/4S_{min}=8\chi^2\bar{n}/\kappa$. This rate coincides with
the total ensemble dephasing (\ref{Gamma-cQED}), and therefore the
qubit state cannot additionally fluctuate due to any other reason.
Thus we derived an important result: {\it in the ideal case with
phase-sensitive amplifier there is only the ``spooky'' back-action
and no ``realistic'' back-action}. This means that the {\it number
of photons in the resonator does not fluctuate} (otherwise there
would be an additional dephasing), which makes sense since we cannot
measure the $A$-quadrature, carrying information on the photon
number. Notice that it is also easy to prove this result when
$\omega_m\neq\omega_r$. Then from Eq.\ (\ref{Fabry-Perot}) we obtain
that the informational quadrature amplitude is multiplied by the
factor $[1+4(\omega_m-\omega_r)^2/\kappa^2]^{-1/2}$ compared with
Eq.\ (\ref{A-B-quadratures}). The response $\Delta I_{max}$ is
multiplied by same factor, while the noise $S_{min}$ does not
change. Therefore, the ``spooky'' back-action contribution into the
ensemble dephasing is multiplied by the factor
$[1+4(\omega_m-\omega_r)^2/\kappa^2]^{-1}$, which again coincides
with the result \cite{Blais-04,Clerk-RMP} for the total ensemble
dephasing $\Gamma$. This proves the absence of the ``realistic''
back-action for the non-resonant case $\omega_m\neq\omega_r$ as
well.

    Now let us consider the case when an ideal phase-sensitive
amplifier amplifies the $A$-quadrature (we again assume
$\omega_m=\omega_r$ for simplicity). Then we do not get any
information on the qubit $z$-coordinate, and therefore there is no
``spooky'' back-action, but there is the ``realistic'' back-action
due to fluctuating number of photons. The description of evolution
in this case is essentially the standard description
\cite{Blais-04,Clerk-RMP}. Let us still denote with $I(t)$ the
output signal from the mixer, corresponding to the amplified
quadrature. For measurement during time $t$ we measure
$A$-quadrature with imprecision $\sigma_{gr}/\sqrt{\kappa t}$. This
is consistent with fluctuation of the number $N$ of emitted photons:
${\rm var}(N)=\bar{N}$, $\bar{N}=\bar{n} \kappa t$. The correlation
function of photon number in the resonator depends on time as $\exp
(-\kappa t/2)$ \cite{Blais-04,Clerk-RMP}, which means that each
extra photon inferred from $I(t)$ fluctuation spends (on average)
time $2/\kappa$ in the resonator and therefore changes the qubit
phase $\phi$ by $4\chi/\kappa$ (the correlation time $2/\kappa$ is
essentially the lifetime of the field, not power
\cite{Blais-04,Clerk-RMP}). Then $\phi$-variance is ${\rm var} (\phi
)=(4\chi/\kappa )^2 \bar{n}\kappa t$, and the corresponding ensemble
dephasing is ${\rm var} (\phi )/2t=8\chi^2\bar{n}/\kappa$. As
expected, this reproduces the standard result (\ref{Gamma-cQED}) for
the ensemble dephasing, while for individual qubit evolution we have
the above discussed correlation: each additional photon inferred
from $I(t)$ fluctuation changes $\phi$ by $4\chi/\kappa$. For the
same amplifier gain and noise as for measuring $B$-quadrature, we
get $\delta \sqrt{\bar{n}}=(4\chi/\kappa) \sqrt{\bar{n}}\,(\delta
I_m)/\Delta I_{max}$, and therefore the correlation is $K=\delta
\langle \phi \rangle/\delta (\overline{I}_m t) = 32
(\chi^2/\kappa)\, \bar{n}/\Delta I_{max}=\Delta I_{max}/S_{min}$. It
is easy to check that $K^2S_{min}/4$ [see Eq.\ (\ref{Gamma})]
coincides with ensemble dephasing $\Gamma$ from Eq.\
(\ref{Gamma-cQED}), as expected for the presence of only
``realistic'' back-action.

    Finally, assume that the phase-sensitive amplifier amplifies the
quadrature, which makes angle $\varphi$ with the optimal
$B$-quadrature and angle $\pi/2-\varphi$ with the $A$-quadrature.
The measured signal $I(t)$ still denotes the output of the IQ mixer,
corresponding to the amplified quadrature;  now it gives information
about both $B$ and $A$ quadratures, with the factors $\cos\varphi$
and $\sin\varphi$, respectively. Combining the ``spooky'' and
``realistic'' back-actions, we get the same formulas as for the
broadband detection of Sec.\ \ref{broadband}:
   \begin{eqnarray}
&&   \frac{\rho_{11}(t)}{\rho_{00}(t)} =
    \frac{\rho_{11}(0)}{\rho_{00}(0)} \, \exp [
 \tilde{I}_m(t) \Delta I/D], \,\,\, D=\frac{S}{2t}, \qquad
    \label{ph-sens-diag}\\
&&  \frac{\rho_{01}(t)}{\rho_{01}(0)}= \frac{\sqrt{\rho_{00}(t)\,
\rho_{11}(t)}} {\sqrt{\rho_{00}(0)\, \rho_{11}(0)}}\, \exp [i K
\tilde{I}_m(t)\, t] ,
    \label{ph-sens-off}\\
&& \Delta I= I_1-I_0= \Delta I_{max}\cos\varphi, \,\,\,
   K=K_{max} \sin\varphi , \qquad
   \label{ph-sens-DeltaI}\\
&&   K_{max} = \Delta I_{max}/S,
     \label{ph-sens-Kmax}\\
&&   \tilde I_m(t) = \frac{1}{t} \int_0^\infty I(t')\, dt'
   - \frac{I_0+I_1}{2},
   \label{ph-sens-Im}\\
&& I(t)=\frac{I_0+I_1}{2} + \frac{\Delta I}{2} \, z (t) +\xi(t),
\,\,\, S_\xi(\omega )=S.
    \label{ph-sens-I(t)}\end{eqnarray}
Here we introduced $\tilde I_m$ by subtracting the constant
$(I_0+I_1)/2$ from $\overline{I}_m$ and did a simple algebra to
convert Eq.\ (\ref{rho-diag}) into Eq.\ (\ref{ph-sens-diag}); the
qubit rotating frame corresponds to $\bar{n}$ photons, $K_{max}$ is
the above discussed correlation for $A$-quadrature amplification,
$\Delta I_{max}$ is the response for $B$-quadrature amplification,
and $S=S_{min}$. Notice that the total ensemble dephasing
(\ref{Gamma}) does not depend on $\varphi$:
    \be
    (\Delta I_{max}\cos\varphi )^2/4S_{min}+
    (K_{max}\sin\varphi )^2 S_{min}/4=  \Gamma.
    \ee

    So far we discussed only the ideal case. There are several
mechanisms for non-ideality. First, the qubit may have additional
environmental dephasing $\gamma_{env}$. This will lead to the extra
factor $e^{-\gamma_{env} t}$ in Eq.\ (\ref{ph-sens-off}) and
increase the ensemble dephasing $\Gamma$ by $\gamma_{env}$.
Following the definitions in Sec.\ \ref{broadband}, the
corresponding quantum efficiency is $\eta_{env}=(1+\gamma_{env}
\kappa/8\chi^2\bar{n})^{-1}$. Second, not all microwave power
leaking from the resonator may be collected and amplified. This can
be characterized by the collection efficiency
$\eta_{col}=\kappa_{col}/\kappa$ and multiplies the response $\Delta
I$ and correlation $K$ by the factor $\sqrt{\eta_{col}}$, while not
affecting the output noise $S$. Third, if the phase-preserving
amplifier is not quantum-limited, it introduces additional noise
$S_{add}$ compared with the quantum limit $S_{min}$ [given by Eq.\
(\ref{S}) when $\eta_{col}=1$]. The corresponding amplifier
efficiency is $\eta_{amp}=S_{min}/(S_{min}+S_{add})$. This does not
affect $\Delta I$ but multiplies $K$ by $\eta_{amp}$ (because for
uncorrelated Gaussian-distributed random numbers $x_1$ and $x_2$,
the averaging of $x_1$ for a fixed sum $x_1+x_2$ gives the
correlation $\langle x_1\rangle/(x_1+x_2)={\rm var}(x_1)/[{\rm
var}(x_1)+{\rm var}(x_2)]$).

If all three mechanisms of the non-ideality are present, then the
evolution can still be described \cite{Kor-nonideal} by Eqs.\
(\ref{ph-sens-diag})--(\ref{ph-sens-I(t)}), but $S$ is now the total
(experimental) output noise, $\Delta I_{max}$ is the experimental
response for $\varphi=0$, so that $S=(\Delta I_{max})^2
\kappa/(32\chi^2\bar{n}\,\eta_{col}\,\eta_{amp})$, the correlation
$K=\delta\langle\phi\rangle /\delta (\tilde{I}_mt)$ is still given
by Eqs.\ (\ref{ph-sens-DeltaI})--(\ref{ph-sens-Kmax}), and the only
change is the extra factor in Eq.\ (\ref{ph-sens-off}):
   \begin{eqnarray}
&& \frac{\rho_{01}(t)}{\rho_{01}(0)}= \frac{\sqrt{\rho_{00}(t)\,
\rho_{11}(t)}} {\sqrt{\rho_{00}(0)\, \rho_{11}(0)}}\, \exp [i K
\tilde{I}_m(t)\, t] \, e^{-\gamma t} , \qquad
    \label{ph-sens-off-2}\\
&& \gamma =\Gamma -(\Delta I_{max})^2/4S,
    \label{ph-sens-gamma}\end{eqnarray}
where the ensemble dephasing is now $\Gamma
=8\chi^2\bar{\eta}/\kappa +\gamma_{env}$. We emphasize that the
qubit evolution depends only on the experimentally measurable
parameters $\Delta I_{max}$, $S$, $\Gamma$, $\varphi$, and the
output signal $I(t)$.

The quantum efficiencies can be expressed as
    \be
\eta=\frac{(\Delta
I_{max})^2}{4S\Gamma}=\eta_{amp}\,\eta_{col}\,\eta_{env} , \,\,\,
\tilde{\eta}=\eta \cos^2\varphi,
    \ee
where as in Sec.\ \ref{broadband}, $\eta$ is the relative
contribution to $\Gamma$ from both the ``spooky'' and ``realistic''
back-actions, while $\tilde\eta$ is the relative contribution from
only the ``spooky'' back-action. The definition of $\tilde\eta$
corresponds to replacing the ``realistic'' back-action factor
$\exp(iK\tilde{I}_m t)$ in Eq.\ (\ref{ph-sens-off-2}) with the
corresponding ensemble dephasing $\exp (-K^2St/4)$. As mentioned in
Sec.\ \ref{broadband}, the peak-to-pedestal ratio of the spectral
peak of contiuous Rabi oscillations is
$4\tilde{\eta}=4\eta\cos^2\varphi$.

    \subsection{Phase-preserving amplifier}

    Now assume that a phase-preserving amplifier is used (this
includes parametric amplifier, HEMT, etc.). Now both the
$A$-quadrature and $B$-quadrature of Eq.\ (\ref{A-B-quadratures})
are amplified independently with the same gain. Correspondingly,
both quadratures at the IQ mixer output carry physical information
instead of only one quadrature in the case of a phase-sensitive
amplifier. Let us denote with $I(t)$ the output of the IQ mixer,
corresponding to the $B$-quadrature; thus $I(t)$ provides
information on the qubit $z$-coordinate. The output signal for the
orthogonal quadrature is denoted $Q(t)$; it corresponds to the
$A$-quadrature in the resonator and provides information on the
fluctuating number of photons. The main difference from the case of
a phase-sensitive amplifier is that now the ``spooky'' and
``realistic'' back-actions are related to two different output
signals: $I(t)$ and $Q(t)$.

    Let us start with the quantum-limited case and assume
an amplifier with the same gain as in the phase-sensitive case, so
that the $I(t)$-channel response is the same as the optimal
phase-sensitive response, $\Delta I=\Delta I_{max}$. The ``spooky''
back-action is always described by the quantum Bayes formulas
(\ref{rho-diag})--(\ref{rho-off}), but now the noise $S$ of the
output $I(t)$ is twice larger than the value (\ref{S}) for the
phase-sensitive amplifier, $S=2S_{min}$ (see discussion in Sec.\
\ref{amplifiers}); therefore the ``spooky'' evolution is twice
slower than in the phase-sensitive case with $\varphi=0$.
 The signal $Q(t)$ has the same noise $S=2S_{min}$; it is again
 twice larger than for the phase-sensitive case with
 $\varphi=\pi/2$; therefore the correlation factor
$K=\delta\langle\phi\rangle/\delta [\int_0^t Q(t')\, dt']$ for the
``realistic'' back-action is twice smaller: $K=K_{max}/2$ (this
reduction is similar to the effect of a non-ideal amplifier
discussed above). We see that $K=\Delta I/S$, and the ensemble
dephasing is at least $(\Delta I)^2/4S+K^2S/4=(\Delta
I)^2/2S=(\Delta I_{max})/4S_{min}$. This again coincides with
$\Gamma=8\chi^2\bar{n}/\kappa$, and therefore there can be no
additional evolution of the qubit besides these ``spooky'' and
``realistic'' back-actions.

 Thus in the ideal case the qubit evolution is
   \begin{eqnarray}
&&   \frac{\rho_{11}(t)}{\rho_{00}(t)} =
    \frac{\rho_{11}(0)}{\rho_{00}(0)} \, \exp [
 \tilde{I}_m(t) \Delta I/D], \,\,\, D=\frac{S}{2t}, \qquad
    \label{ph-pres-diag}\\
&&  \frac{\rho_{01}(t)}{\rho_{01}(0)}= \frac{\sqrt{\rho_{00}(t)\,
\rho_{11}(t)}} {\sqrt{\rho_{00}(0)\, \rho_{11}(0)}}\, \exp [i K
\tilde{Q}_m(t)\, t] ,
    \label{ph-pres-off}\\
&& \Delta I= I_1-I_0, \,\,\,  K = \Delta I/S,
     \label{ph-pres-K}\\
&&   \tilde Q_m(t) = \frac{1}{t} \int_0^\infty Q(t')\, dt'
   - \langle Q\rangle, \,\,\, S_Q=S_I=S,
     \label{ph-pres-Q}\end{eqnarray}
where $\langle Q\rangle$ is the average value of $Q(t)$ (which
depends on $\bar{n}$), $\tilde{I}_m(t)$ is defined by Eq.\
(\ref{ph-sens-Im}), and the channels $I(t)$ and $Q(t)$ both have the
same (uncorrelated) noise $S=(\Delta I)^2\kappa/(16\chi^2\bar{n})$.
Notice that $(\Delta I)^2/4S=K^2S/4=4\chi^2\bar{n}/\kappa$, and
therefore {\it in the phase-preserving case the ensemble dephasing
$\Gamma$ contains equal contributions $\Gamma/2$ from the ``spooky''
and ``realistic'' back-actions}. We emphasize that Eqs.\
(\ref{ph-pres-diag})--(\ref{ph-pres-off}) still {\it allow us to
monitor a qubit wavefunction} if the initial qubit state is pure.

    A non-ideal case can be analyzed in the same way as for the
phase-sensitive amplifier. An extra dephasing $\gamma_{env}$ of the
qubit is described by
$\eta_{env}=(1+\gamma_{env}\kappa/8\chi^2\bar{n})^{-1}$, imperfect
collection efficiency is described by
$\eta_{col}=\kappa_{col}/\kappa$, and the amplifier efficiency is
$\eta_{amp}$. We define $\eta_{amp}=S_{ql}/S$ for a phase-preserving
amplifier by comparing its output noise $S$ (per quadrature) with
the quantum limit for a phase-preserving amplifier:
$S_{ql}=2S_{min}$, so that $\eta_{amp}=1$ in the quantum-limited
case. We also define $\tilde\eta=S_{min}/S$ by comparison with a
phase-sensitive amplifier having the same gain, so that
$\tilde{\eta}_{amp}=\eta_{amp}/2$ and obviously
$\tilde{\eta}_{amp}\leq 1/2$. Similarly to the phase-preserving
case, incomplete microwave collection multiplies the response
$\Delta I$ and correlation $K$ by the factor $\sqrt{\eta_{col}}$ but
does not change the noise $S$; the extra noise in the amplifier
multiplies $K$ by $\eta_{amp}$ but does not change $\Delta I$.

The qubit evolution can still be described by Eqs.\
(\ref{ph-pres-diag})--(\ref{ph-pres-Q}) with the only change in Eq.\
(\ref{ph-pres-off}):
    \begin{eqnarray}
&& \frac{\rho_{01}(t)}{\rho_{01}(0)}= \frac{\sqrt{\rho_{00}(t)\,
\rho_{11}(t)}} {\sqrt{\rho_{00}(0)\, \rho_{11}(0)}}\, \exp [i K
\tilde{Q}_m(t)\, t] \, e^{-\gamma t} , \qquad
    \label{ph-pres-off-2}\\
&& \gamma=\Gamma-2(\Delta I)^2/4S,
    \label{ph-pres-gamma}\end{eqnarray}
where now $S$ is the total (experimental) noise per quadrature,
$\Delta I$ is the experimental response, and $\Gamma
=8\chi^2\bar{\eta}/\kappa +\gamma_{env}$ is the total ensemble
dephasing. The qubit evolution is determined by the parameters
$\Delta I$, $S$, $\Gamma$, and output signals $I(t)$ and $Q(t)$.

    The quantum efficiencies are
    \be
 \eta=\eta_{amp}\eta_{col}\eta_{env}=(\Delta I)^2/2S\Gamma, \,\,\,
\tilde\eta =\eta/2.
    \ee
Here $\eta$ is the fraction of $\Gamma$ due to the contribution from
both the ``spooky'' and ``realistic'' back-actions.  The efficiency
$\tilde\eta =\tilde{\eta}_{amp}\eta_{col}\eta_{env}$ is the fraction
from only the ``spooky'' contribution; it corresponds to replacing
the term $\exp (i K \tilde{Q}_m\, t)$ in Eq.\ (\ref{ph-pres-off-2})
with the dephasing term $\exp(-K^2t/4S)$. In particular, the
peak-to-pedestal ratio of the Rabi spectral peak for the signal
$I(t)$ is $4\tilde\eta=2\eta$.

    Let us mention that Eqs.\
(\ref{ph-pres-diag})--(\ref{ph-pres-K}) for the ideal
phase-preserving case can also be obtained from Eqs.\
(\ref{ph-sens-diag})--(\ref{ph-sens-Kmax}) for the phase-sensitive
case in the following way.
   Let us think about a phase-preserving
amplifier as a phase-sensitive amplifier, in which the angle
$\varphi$ rapidly changes with time, and we have to average over
$\varphi$. When coefficients $\cos\varphi$ and $\sin\varphi$ in Eq.\
(\ref{ph-sens-DeltaI}) are substituted into Eqs.\
(\ref{ph-sens-diag}) and (\ref{ph-sens-off}), we see a natural
formation of the quadratures $\tilde{I}_m$ and $\tilde{Q}_m$ of the
phase-preserving setup. Then the exponential factor in Eq.\
(\ref{ph-sens-diag})  becomes $\exp (\tilde{I}_m \Delta I_{max}/D)$,
and the exponential factor in Eq.\ (\ref{ph-sens-off})  becomes
$\exp (i K_{max}\tilde{Q}_m t)$. Now let us take into account that
the average response is $\Delta I=\overline{\cos^2\varphi}\, \Delta
I_{max}=\Delta I_{max}/2$, and the phase-sensitive amplifier noise
$S$ splits equally between the $I(t)$ and $Q(t)$ quadratures (the
orthogonal, de-amplified quadrature is noiseless). The mutual
cancellation of these two factors of 2 leads to the same form of
Eq.\ (\ref{ph-pres-diag}) as in Eq.\ (\ref{ph-sens-diag}) and the
relation $K=\Delta I/S$ in Eq.\ (\ref{ph-pres-K}).

    One more way to understand the relation between the ideal
phase-sensitive and phase-preserving cases is the following. Instead
of using a phase-preserving amplifier, let us split the outgoing
microwave into two equal parts and use phase-sensitive amplifiers
with $\varphi=0$ and $\varphi=\pi/2$ in the two channels. To keep
the same noise $S$ per channel, we increase the gain by the factor
$\sqrt{2}$, that also compensates the signal loss at the splitter.
Then the channel $\varphi=0$ produces the ``spooky'' back-action
(\ref{ph-pres-diag}), while the channel $\varphi=\pi/2$ informs us
of the ``realistic'' back-action (\ref{ph-pres-off}), and the
relation (\ref{ph-pres-K}) between $K$ and $\Delta I$ is the same as
between $K_{max}$ and $\Delta I_{max}$ in (\ref{ph-sens-Kmax}).

    In the ideal case ($\eta=1$) the qubit evolution description
can be translated into the language of the POVM-type measurement. In
the same way as in Sec.\ \ref{broadband}, Eqs.\ (\ref{ph-pres-diag})
and (\ref{ph-pres-off}) can be converted into the measurement
operator
    \begin{eqnarray}
&&    M(\overline{I}_m,\tilde Q_m) = \exp ( -i K \tilde{Q}_m t \,
\sigma_z/2)
    \nonumber \\
&&\hspace{0.5cm} \times \left[\sqrt{P_{|0\rangle}(\overline{I}_m)}
\, |0\rangle \langle 0| + \sqrt{P_{|1\rangle}(\overline{I}_m)} \,
|1\rangle \langle 1|\right], \qquad
    \label{meas-op-QED}\end{eqnarray}
where the probabilities $P_{|j\rangle}$ are given by Eq.\
(\ref{gaussian}). Similarly, Eqs.\ (\ref{ph-sens-diag}) and
(\ref{ph-sens-off}) for the case of a phase-sensitive amplifier can
be converted into the same measurement operator (\ref{meas-op-QED}),
in which $\tilde Q_m$ is replaced with $\tilde I_m$.

\section{Conclusion}

    We have presented a simple physical picture of the qubit
evolution due to its measurement in the circuit QED setup. The
``spooky'' back-action is universal, it is caused by gradual
extraction of information about the qubit state. The ``realistic''
back-action is due to a specific mechanism: fluctuation of the
photon number in the resonator. For a phase-sensitive amplifier the
qubit evolution is described by Eqs.\ (\ref{ph-sens-diag}) and
(\ref{ph-sens-off-2}); it is determined by the output signal $I(t)$,
which corresponds to the amplified quadrature. For a
phase-preserving amplifier the evolution is described by Eqs.\
(\ref{ph-pres-diag}) and (\ref{ph-pres-off-2}); it is determined by
two output signals: $I(t)$ and $Q(t)$, where $I(t)$ now corresponds
to the quadrature, which provides information about the qubit state
($B$-quadrature in the resonator) and $Q(t)$ corresponds to the
orthogonal $A$-quadrature, which gives us a record of the photon
number fluctuations in the resonator.

    While the cQED setup significantly differs from both the
broadband quantum measurement setup \cite{Kor-rev} and the standard
optical setup \cite{quant-traj}, we see that the description of the
qubit evolution is exactly the same as in both these cases if a
phase-sensitive amplifier is used. The description is only slightly
different when a phase-preserving amplifier is used: we should
assign the ``spooky'' and ``realistic'' back-actions to the separate
output signals $I(t)$ and $Q(t)$ instead of only one signal. It is
also useful to think about the phase-preserving case via the model
in which we split the outgoing microwave (the quantum signal) into
two equal parts and then use 90-degree-shifted phase-sensitive
amplifiers for these two channels.

    We intentionally considered only the simplest case, because most
of the further steps and generalizations are quite straightforward
\cite{Kor-rev}. In particular, it it very simple to include Rabi
oscillations and energy relaxation of the qubit state. For that we
have to take time-derivative of the evolution equations and add the
terms due to Rabi oscillations and energy relaxation. If the
Stratonovich definition of the derivative is used, we get the
equations of the Bayesian formalism \cite{Kor-rev}; if the It\^o
derivative is used, we get the equations of the quantum trajectory
formalism \cite{quant-traj,Gambetta-08}. Generalization to
measurement of several entangled qubits is also straightforward
\cite{Kor-rev}. We have considered only the resonant case,
$\omega_m=\omega_r$; however, generalization to the case
$\omega_m\neq \omega_r$ is quite simple (see \cite{Gambetta-08}): we
just need a different definition of the informational $B$-quadrature
and photon-fluctuation $A$-quadrature. In our formalism we
implicitly assumed sufficiently wide bandwidth of the amplifier
(much larger than the ensemble dephasing $\Gamma$ and Rabi frequency
$\Omega_R$). If this is not the case, the formalism should change
significantly. However, we believe that in most of the practical
cases we can take this effect into account by adding a classical
narrowband filter to the classical signal at the amplifier output;
this will correspond to passing the signals $I(t)$ and $Q(t)$
through the low-pass filters. A much more serious change of the
theory is required when the resonator bandwidth $\kappa$ is
comparable to $\Gamma$ or $\Omega_R$; this still has to be done.

    Understanding the difference between the ``spooky'' and
``realistic'' back-actions is important for designing the quantum
feedback control of the Rabi oscillations \cite{Kor-feedback}. The
simplest case is when a phase-sensitive amplifier amplifies the
informational $B$-quadrature. Then there is no ``realistic''
back-action, and the feedback loop should only modulate the
amplitude of the Rabi drive (i.e.\ the Rabi frequency $\Omega_R$);
this case was well studied for the broadband setup
\cite{Kor-feedback}. The situation is different for a
phase-preserving amplifier. Then we need two feedback channels: the
first (usual) channel should modulate the Rabi frequency $\Omega_R$
to compensate the ``spooky'' back-action, while the second channel
should compensate the ``realistic'' back-action by modulating the
qubit frequency $\omega_{qb}$ or the the frequency of the Rabi drive
$\omega_R$. The controller for the second feedback channel is quite
simple: it should compensate the contribution $iK\tilde{Q}(t)$ to
the qubit phase derivative $\dot{\phi}(t)$ due to the $K$-term in
Eq.\ (\ref{ph-pres-off-2}). Therefore the controller is
    \be \Delta(\omega_{qb}-\omega_{R})=
    -K[Q(t)-\langle Q\rangle],
    \ee
i.e. we should directly apply the signal $Q(t)$ to modulate
$\omega_{qb}$ or $\omega_R$. The second feedback channel essentially
eliminates the $K$-term in Eq.\ (\ref{ph-pres-off-2}) and decreases
the ensemble dephasing $\Gamma$ by $K^2S/4=(\Delta I)^2/4S$.
Correspondingly, in the absence of the first (main) feedback channel
the peak-to-pedestal ratio of the Rabi peak increases from 2 to 4 in
the quantum-limited case. The first feedback channel should be the
same as for the broadband setup; it depends on the signal $I(t)$ and
can be realized using various ideas for the controller (``direct'',
Bayesian, ``simple'', etc. \cite{Kor-feedback}). Notice that {\it
without the second channel the feedback performance is determined by
the quantum efficiency $\tilde\eta$, while with the second channel
it is determined by
$\tilde{\tilde\eta}=\tilde\eta/(1-\eta+\tilde\eta)$ } (this is one
more combination of the terms in Eq.\ (\ref{Gamma}), which can be
used for the definition \cite{Kor-rev} of quantum efficiency). The
case of a phase-sensitive amplifier, which amplifies a non-optimal
quadrature ($\varphi\neq 0$, $\varphi\neq \pi/2$) is similar to the
case of a phase-preserving amplifier, but both feedback channels
should start with the same signal $I(t)$. {\it In both the
phase-sensitive and phase-preserving setups a perfect feedback
control is possible in the quantum-limited case $\eta=1$.}

    Discussion of the ``spooky'' and ``realistic'' back-actions in
the cQED setup necessarily raises the question of causality. When
the microwave leaves the resonator, it does not yet ``know'' in
which way it will be measured (phase-preserving or phase-sensitive,
which angle $\varphi$, etc.). Moreover, when a circulator is used
for the outgoing microwave, the field in the resonator and the qubit
can never ``know'' in a realistic way which method of measurement is
used. Nevertheless, the qubit evolution strongly depends on the
measurement method. As we discussed, the ``spooky'' evolution moves
the qubit state along the meridians of the Bloch sphere, the
``realistic'' back-action moves the state along the parallels, and
the measurement method determines whether the qubit experiences the
``spooky'' or ``realistic'' back-action (or their combination). In
this sense the ``realistic'' back-action is not fully realistic: it
has the physical mechanism, but whether this mechanism works or not
is determined in a spooky way. The causality requires that we cannot
pass a ``useful'' information to the qubit by choosing the
measurement method. This means that {\it the ensemble-averaged
evolution of the qubit cannot depend on the measurement method (this
is the general requirement of causality in quantum mechanics)}. It
is surely satisfied in our cQED setup.

    The author thanks Michel Devoret, Konstantin Likharev, Patrice
Bertet, and  Farid Khalili for useful discussions. The work was
supported by ARO MURI grant W911NF-11-1-0268 and by NSA/IARPA/ARO
grant W911NF-10-1-0334.



\begin{thebibliography}{99}
\bibitem{quadrature} We use the word ``quadrature'' in a somewhat
sloppy way.
\end{thebibliography}
\end{document}